\documentclass[]{jacow}

\usepackage[english]{babel}
\usepackage{bm}
\usepackage{booktabs}
\usepackage{graphicx}
\usepackage{siunitx}

\newcommand{\bestn}{Best-$N$}
\ifPDFTeX
  \newcommand{\tablemathbold}[1]{\bm{#1}}
\else
  \newcommand{\tablemathbold}[1]{\symbf{#1}}
\fi
\newcommand{\PrimaryHBestOneScore}{0.343}

\newcommand{\PrimaryHBestFiveScore}{0.403}
\newcommand{\PrimaryVBestOneScore}{0.420}

\newcommand{\PrimaryVBestFiveScore}{0.524}

\newcommand{\BestOneUniqueH}{60}
\newcommand{\BestOneUniqueV}{60}
\newcommand{\BestOneMaxFrequencyH}{3.7}
\newcommand{\BestOneMaxFrequencyV}{5.7}
\newcommand{\RidgeHIqrDeltaMilli}{-2.43}
\newcommand{\RidgeHIqrLowMilli}{-3.08}
\newcommand{\RidgeHIqrHighMilli}{-1.93}
\newcommand{\RidgeVIqrDeltaMilli}{1.02}
\newcommand{\RidgeVIqrLowMilli}{0.39}
\newcommand{\RidgeVIqrHighMilli}{1.65}
\newcommand{\BestNCurveSpillPlaneCount}{4000}
\newcommand{\BestNValidationSpillPlaneCount}{1000}
\newcommand{\BestNDigitizerFoldCount}{5}
\newcommand{\BestNHSensitivityAvailable}{5}

\newcommand{\BestNHSensitivityMinimum}{2}
\newcommand{\BestNHSensitivityMaximum}{13}
\newcommand{\BestNVSensitivityAvailable}{6}

\newcommand{\BestNVSensitivityMinimum}{10}
\newcommand{\BestNVSensitivityMaximum}{28}
\newcommand{\AllTrainingHSelectedFavored}{2}
\newcommand{\AllTrainingHBaselineFavored}{3}

\newcommand{\AllTrainingVSelectedFavored}{3}
\newcommand{\AllTrainingVBaselineFavored}{3}

\newcommand{\PrimarySpillCount}{2000}
\newcommand{\PrimaryNominalHChannels}{60}
\newcommand{\PrimaryNominalVChannels}{60}
\newcommand{\PrimaryPartialCaptures}{12}
\newcommand{\PrimarySourceAbsences}{16}

\begin{document}

\title{Turn-by-turn tune analysis using adaptive \NoCaseChange{BPM} ensembles in the Fermilab \NoCaseChange{Mu2e} Delivery Ring\thanks{This manuscript has been authored by Fermi Forward Discovery Group, LLC under Contract No. 89243024CSC000002 with the U.S. Department of Energy, Office of Science, Office of High Energy Physics.}}

\author{D. Steinkamp\thanks{derekste@fnal.gov}}
\affil{Fermi National Accelerator Laboratory, Batavia, IL, USA}

\maketitle

\begin{abstract}
The Mu2e experiment at Fermilab requires stable resonant slow extraction from the Delivery Ring, making reliable tune monitoring an important operational diagnostic. This work investigates BPM-based tune-candidate extraction using synchronized turn-by-turn position data distributed across multiple digitizers. Each spill contains approximately 50,000 turns from many BPMs in both transverse planes, enabling spectral analysis of tune-like structure.

The analysis captures coherent spill snapshots, verifies synchronization using stream timestamps, and computes tune candidates in configurable horizontal and vertical tune bands. Rather than relying on a single BPM or fixed BPM list, it evaluates BPM quality on a spill-by-spill basis and selects small adaptive BPM ensembles. A multi-spill study shows that tune observability is distributed and dynamic rather than concentrated in one globally optimal BPM. Adaptive ensembles improve tune-candidate quality compared with single-BPM selections, with the clearest results in the vertical plane. The horizontal plane shows useful ranking structure but weaker visibility under present thresholds.

Direct evaluation of fixed global BPM sets shows that static selections do not reproduce dynamic per-spill performance. These results motivate an adaptive BPM-ensemble approach for Delivery Ring tune analysis using selected BPM subsets, confidence metrics, and quality flags rather than a single preferred BPM or fixed BPM list.
\end{abstract}

\section{Introduction}

The Mu2e slow-extraction programme uses third-integer resonant extraction from the Fermilab Delivery Ring. Spill regulation depends on a tune-ramp quadrupole family and radio-frequency knock-out excitation, making transverse tune observability relevant to both commissioning and operations~\cite{ibrahim:ibic2019-mopp033,nagaslaev:prab2019,nagaslaev:commissioning2026}. The installed Muon Campus BPM system provides synchronized turn-by-turn (TBT) position waveforms throughout the ring~\cite{patel:ibic2019-wepp044}. A single BPM, however, need not remain the best observer as orbit, oscillation amplitude, extraction conditions, and instrument quality change.

Combining TBT data from multiple BPMs can improve tune estimation when individual signals have unequal signal-to-noise ratio or decohere rapidly~\cite{zisopoulos:ipac2017-mopab122,alexahin:ipac2010-mope084}. The central question here is therefore not which BPM is globally best, but how many BPMs should be combined for each spill and whether that adaptive choice is reproducible outside the data used to make it.

This paper reports a BPM-only internal validation. The reported values are fractional tune candidates in configured search bands. No Schottky, tune-meter, or controlled tune-knob reference was available for these captures, so the analysis does not claim absolute tune accuracy.

\section{Data and synchronization}

The primary data set contains \PrimarySpillCount{} captured spills in two acquisition collections. Its nominal topology supplies \PrimaryNominalHChannels{} H and \PrimaryNominalVChannels{} V position channels, with two same-plane sources per digitizer. A fresh raw-payload audit records \PrimaryPartialCaptures{} partial captures and \PrimarySourceAbsences{} source absences; missing channels remain missing rather than being zero-filled. Each waveform contains approximately 50,000 turns. Stream timestamps assemble the spill snapshot: adjacent target buckets merge within \qty{1}{ms}, and each captured stream must satisfy a separate \qty{25}{ms} same-spill tolerance. The host-published ``raw'' array is a DSP-derived position before threshold substitution and scaling, not an electrode ADC waveform.

Publication auditing exposed two identity defects in an earlier analysis: a shared digitizer label had replaced the exact channel key, and ring order had been parsed from the wrong address field. The corrected pipeline records exact source identities and resolves every subset bit mask before labels are displayed; detailed provenance and regression tests are retained with the repository.

The waveform path subtracts the window mean, applies a Hann taper, and computes a zero-padded real fast Fourier transform. Frequencies are mapped to the upper fractional tune branch by $q=1-f$. The discovery bands are $0.60\leq q_x\leq0.70$ and $0.67\leq q_y\leq0.75$; expected values near 0.65 and 0.72 enter only as soft priors.

\section{Adaptive ensemble method}

For the primary search, overlapping 4,096-turn windows with 256-turn stride characterize early-spill spectral structure. Per-BPM peak candidates are clustered within each spill and plane to form an internal consensus; no distribution across other spills is used as a label. Candidate subset spectra are averaged in power. In plain terms, one part of the data chooses the BPM members and candidate tune, while later non-overlapping data test whether that choice persists. Turn-dependent diagnostics are retained because extraction records are nonstationary~\cite{russo:prab2024}.

The recorded subset score can be written compactly as
\begin{align}
s_w &= .35U_w+.2L_w+.15C_w+.1(V_w+D_w+F_w)-A_w,\nonumber\\
S &= .6\,\mathrm{med}_w\,s_w+.25\,P_{10,w}(s_w)+.15W.\nonumber
\end{align}
Here $U$ is nonselected training-channel support, $L$ combines prominence and spectral entropy, $C$ is within-spill consensus agreement, $V$ is visibility, $D$ is digitizer/ring diversity, $F$ is finite-window quality, $A$ is ambiguity, and $W$ is window stability. Thus $S$ ranks subsets; it is not a measurement uncertainty.

Best-1 and Best-3 are searched exhaustively over the 60 channels in a plane. Best-5 uses exhaustive enumeration within a recorded 20-channel screening pool. The diversity-independent Best-1 result directly shows changing observability: all \BestOneUniqueH{} H and \BestOneUniqueV{} V sources win at least once, while the most frequent winners account for only \BestOneMaxFrequencyH{}\% and \BestOneMaxFrequencyV{}\% of spills, respectively.

\subsection{Leakage-Controlled \bestn{} Selection}

The ensemble-size study sweeps contiguous $N$ values instead of comparing only 1, 3, and 5. Members are selected from eight fit windows. Every later window overlapping that prefix is purged, and evaluation begins with the first non-overlapping window. Expected-tune priors and precomputed consensus are disabled during this Best-$N$ validation; candidate tunes are derived from the training fit windows only.

The full $N$ curve uses \BestNCurveSpillPlaneCount{} spill-plane cases; digitizer-disjoint validation uses an evenly stratified \BestNValidationSpillPlaneCount{}-case subset, each evaluated across \BestNDigitizerFoldCount{} folds rather than being presented as the full curve population.

A bounded beam search is used above $N=1$ and repeated at several beam widths. For channel-disjoint validation, complete digitizers are assigned to five folds so sibling channels never appear on opposite sides. Each fold selects members from training digitizers; later selected-channel spectra are then compared with a median-power reference formed only from held-out digitizers.

The conditioned measure asks whether later spectra support the fitted candidate; the blind measure searches the full band independently on the two channel groups. Agreement means $|\Delta q|\leq0.0025$. The declared operating point is the earliest $N$ passing a 0.02 blind-agreement non-inferiority margin, 95\% selected-power support, 90\% held-out-power support, and the prominence and tune-difference gates. Intervals collapse folds within each spill, then use 1,000 moving-block draws with 20-spill blocks inside each acquisition collection. A deterministic cross-spill block permutation supplies the blind-agreement null. Finally, $N$ is chosen in one collection and transferred to the other; only spill-level membership remains adaptive.

\section{Results}

\subsection{Corrected Adaptive Search}

Across 4,000 corrected spill-plane rows per subset size, median in-sample scores rise from \PrimaryHBestOneScore{} at H Best-1 to \PrimaryHBestFiveScore{} at H Best-5, and from \PrimaryVBestOneScore{} to \PrimaryVBestFiveScore{} vertically. These ranking gains are descriptive; the leakage-controlled later-window results in Fig.~1 and Table~\ref{tab:results} carry the claim.

Seven reduced-sample checks varying beam width, fit-window count, and fold assignment place eligible H choices at Best-\BestNHSensitivityMinimum{}--Best-\BestNHSensitivityMaximum{} in \BestNHSensitivityAvailable{}/7 runs and V choices at Best-\BestNVSensitivityMinimum{}--Best-\BestNVSensitivityMaximum{} in \BestNVSensitivityAvailable{}/7. Gate-margin checks also move the earliest eligible $N$. H Best-5 and V Best-12 are therefore operating points, not unique optima. Score weights, the 4,096-turn spectral-window length, and the $|\Delta q|$ agreement tolerance remain untested sensitivities.

\begin{figure*}[!htb]
  \centering
  \includegraphics[width=0.98\textwidth]{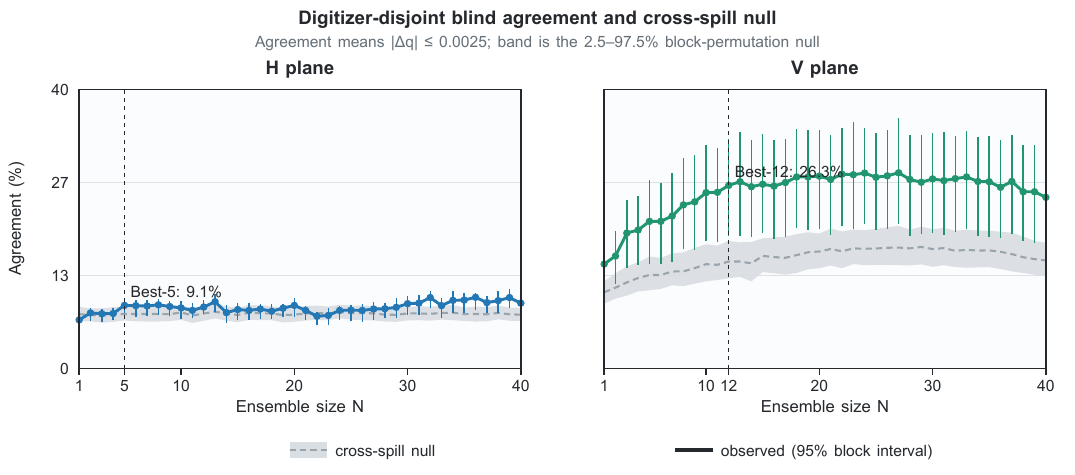}
  \caption{Leakage-controlled \bestn{} validation for H (left) and V (right). Blind full-band selected-versus-held-out-digitizer agreement uses $|\Delta q|\leq0.0025$; points show observed rates and intervals, while the shaded band is the 2.5--97.5 percentile cross-spill null.}
  \label{fig:bestn}
\end{figure*}

\begin{table*}[!htb]
  \centering
  \caption{Leakage-controlled Best-$N$ intervals use collection-preserving spill blocks. Ridge intervals use overlapping-turn blocks on exact-paired picks and describe concentration, not absolute tune accuracy or measured physical noise.}
  \label{tab:results}
  \small
  \begin{tabular}{@{}lcccc@{}}
    \toprule
    \textbf{Plane} & \textbf{Best-$\tablemathbold{N}$} & \textbf{Blind Agreement (\%)} & \textbf{Blind $\tablemathbold{|\Delta q|}$ ($\tablemathbold{10^{-3}}$)} & \textbf{$\tablemathbold{\Delta}$IQR vs B1 ($\tablemathbold{10^{-3}}$)} \\
    \midrule
    H & 5 & 9.1 [7.9, 9.8] & 16.1 [15.5, 17.2] & -2.4 [-3.1, -1.9] \\
    V & 12 & 26.3 [19.1, 33.0] & 10.7 [9.7, 12.0] & 1.0 [0.4, 1.7] \\
    \bottomrule
  \end{tabular}
\end{table*}

\subsection{Full-Spill Ridge-Density Comparison}

To visualize persistence through the approximately 50,000-turn record, the selected early-window membership is fixed and applied to the raw spill. Each trace is RMS-normalized, 4,096-turn Hann spectra are computed every 256 turns, and subset powers are averaged. One locally tracked ridge pick per spill and window is binned across spills; this is a ridge-location density, not a spectral-power heat map. Blank and bounded edge picks remain missing; duplicate traces beneath the methods show the shared exact-paired count by turn.

The primary comparison (Fig.\,2) is corrected adaptive Best-1 against H Best-5 or V Best-12 on exact common spill/window points. One shared scale is used within each plane. This isolates ensemble size from the historical selector repair; the legacy comparison is retained only in the audit gallery. The associated ridge-width contrast (Fig.~3) uses moving-block intervals across turn centers describe whether concentration differences persist despite overlapping windows, not uncertainty over the spill population.

\begin{figure*}[!htb]
  \centering
  \includegraphics[width=0.93\textwidth]{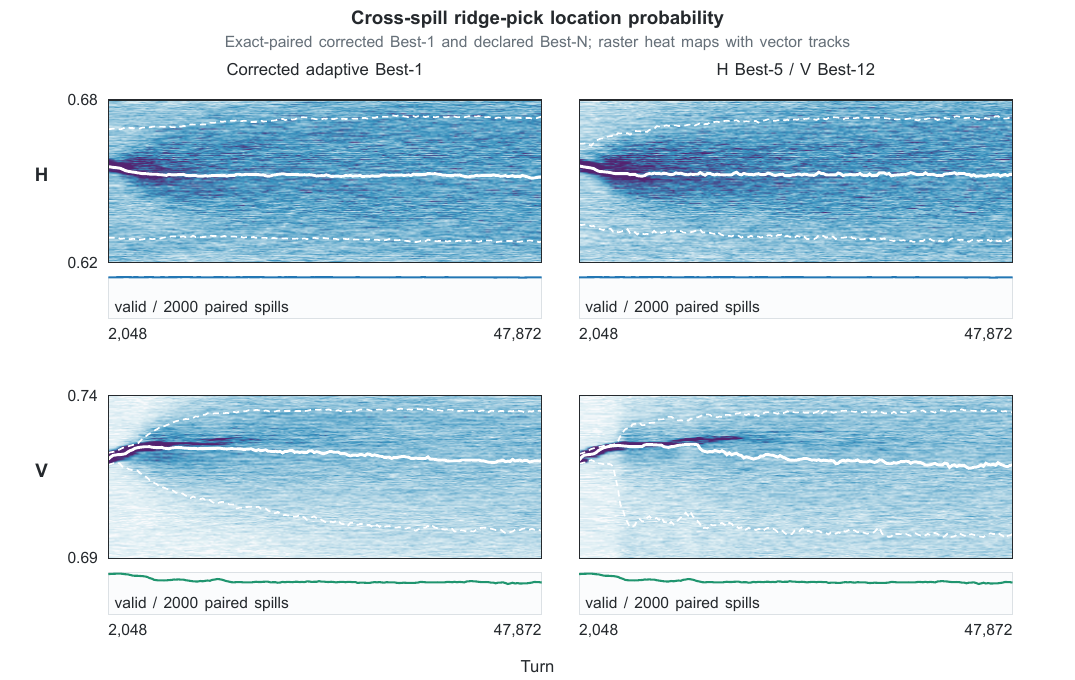}
  \caption{Corrected adaptive Best-1 versus H Best-5/V Best-12 on exact-paired ridge picks. Each panel spans turns 2,048--47,872. Color is column-normalized pick probability; white curves are cross-spill P10/median/P90. Repeated thin traces show the shared valid population.}
  \label{fig:ridge}
\end{figure*}

\begin{figure*}[!htb]
  \centering
  \includegraphics[width=0.82\textwidth]{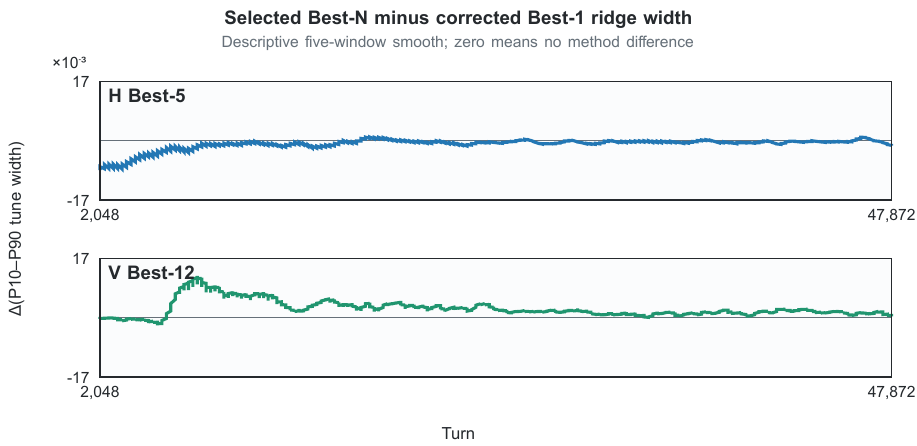}
  \caption{Descriptive exact-paired selected-Best-$N$ minus corrected-Best-1 P10--P90 ridge width. Aggregate intervals are H $\RidgeHIqrDeltaMilli\;[\RidgeHIqrLowMilli,\RidgeHIqrHighMilli]\times10^{-3}$ and V $\RidgeVIqrDeltaMilli\;[\RidgeVIqrLowMilli,\RidgeVIqrHighMilli]\times10^{-3}$. Curves are five-window visual smooths of unsmoothed exports; no pointwise confidence band or physical-noise claim is implied.}
  \label{fig:ridgewidth}
\end{figure*}

\subsection{Controls and Plane Asymmetry}

Frozen global top-1, top-3, and top-5 sets are trained on one collection and applied to the other. Dynamic, frozen, and all-BPM methods are then rescored from the same cached spectra. This control is descriptive because the dynamic memberships reuse their selection windows; the leakage-controlled Best-$N$ test remains the independent result.

Dynamic Best-5 exceeds frozen Best-5 vertically under this descriptive score, but all-training aggregation remains competitive. Under the same purged windows and digitizer folds as Best-$N$, selected subsets are favored in only \AllTrainingHSelectedFavored{} H and \AllTrainingVSelectedFavored{} V comparisons, while all-training mean or median is favored in \AllTrainingHBaselineFavored{} H and \AllTrainingVBaselineFavored{} V; the remainder are unresolved. Adaptation therefore adds auditable membership and fault handling, not universal numerical superiority.

The diagnostics are metric-dependent: V Best-12 gives stronger digitizer-disjoint agreement, whereas H Best-5 clearly narrows the corrected-Best-1 ridge. Blind agreement is low; higher conditioned agreement and below-threshold later visibility are consistent with weak competing peaks, but not their physical cause. H visibility becomes diffuse earlier, so no extraction-onset turn is inferred.

\section{Limitations and operational use}

The study has four principal limits. First, internal BPM agreement is not an independent tune calibration. H Best-5 only just clears its pointwise cross-spill null band, whereas V Best-12 is clearly separated; because $N$ is selected from the same validation curve, the null is diagnostic rather than a post-selection $p$-value. \mbox{Second, unlabeled machine states} limit block-aware uncertainty; controlled scans remain necessary. Third, fixed early-window membership makes the 50,000-turn visualization a persistence test, not full-spill reselection. Fourth, ridge concentration can improve if all methods follow the same biased candidate. A matched Schottky or tune-meter reference and a controlled quadrupole scan comparing measured with optics-predicted $\Delta q$ are therefore required.

Within those limits, adaptive ensembles are useful when channel identity and quality diagnostics matter. All-training aggregation remains an explicit same-protocol control. An operational monitor should expose exact members, ensemble size, tune candidate, prominence and ambiguity, later-window support, and a no-reliable-tune state, with independent decisions in the two planes. The demonstrated workflow is offline and post-spill; an initial deployment should report once after each completed spill. Intra-spill updates require incremental data transport plus separate causal and latency validation.

\section{Conclusion}

Adaptive BPM ensembles recover repeatable tune-like structure. Leakage-controlled tests support H Best-5 and V Best-12 as operating points rather than unique optima. Vertical agreement is stronger across digitizers, horizontal ridge concentration improves relative to corrected Best-1, and all-training aggregation remains competitive. This BPM-only candidate monitor still requires external tune calibration.

\end{document}